# ALTERNATIVES TO SPEECH IN LOW BIT RATE COMMUNICATION SYSTEMS

*Cristina Videira Lopes*
Xerox PARC, Palo Alto, CA
(lopes@parc.xerox.com)

*Pedro M. Q. Aguiar*
ISR-IST, Lisboa, Portugal
(aguiar@isr.ist.utl.pt)

**ABSTRACT**

This paper describes a framework and a method with which speech communication can be analyzed. The framework consists of a set of low bit rate, short-range acoustic communication systems, such as speech, but that are quite different from speech. The method is to systematically compare these systems according to different objective functions such as data rate, computational overhead, psychoacoustic effects and semantics. One goal of this study is to better understand the nature of human communication. Another goal is to identify acoustic communication systems that are more efficient than human speech for some specific purposes.

## 1. INTRODUCTION

Experiments have shown that at the highest levels of the speech process the information rate of human communications is fairly low, less than 100 bps. Speech, without artificial amplification, has a limited range of a few tens of meters and sharp signal declines in the presence of obstacles such as walls. These characteristics make speech seem pre-historical when compared to the modern digital communication systems, which can achieve bit rates in the order of megabits per second and can communicate over very large distances. Speech, however, though slow and limited in range, is a remarkable communication system with characteristics that may also be desired in artificial systems.

Inspired by ubiquity of speech, we have been designing and implementing air modems using ordinary microphones and speakers found in most computers and PDAs. Air modems have never been taken seriously, for two good reasons. First, the data rate of air modems is much lower than those achieved over radio or electric wires, especially if we want to use ordinary microphones and speakers. Second, air modems can be annoying to humans. However, as more and more small, mobile devices support a channel for voice and music, air becomes a cheap option for transferring small amounts of information between devices that happen to be near each other. However, the criteria for designing such communication systems must account for the specific characteristics of the aerial channels. These characteristics impose an interesting set of constraints, some of which, we note, are also constraints for human acoustic communication.

A recent study of air modems can be found in the Gerasimov and Bender's paper [2]. They evaluate standard modulations according to the bit rate, computational overhead, noise tolerance and disruption level. They report a maximum bit rate of 3.4 Kbps going into the low ultrasound band (18KHz). There has also been some work in non-verbal human-computer interactions, see [3]. The goal of that work is to reduce the computation overhead of the voice recognizer by avoiding the task of identifying words.

Our work looks at low bit rate communication systems under a new perspective. We call each of the air modems a "digital voice," and we believe they can be evaluated and compared in a framework of communication systems that also includes human acoustic communication. By varying certain design parameters we can measure how well the signals perform under certain objective functions. The comparison method used in this paper is an important step towards a better understanding of human communication, and one that hasn't been explored before.

The paper is organized as follows. In section 2 we revisit the issue of the bit rate of human speech. Section 3 overviews the approach taken in the Digital Voices project, and describes the channel used, comparing it to the human acoustic channel. Section 4 details several digital voices we have designed. Section 5 presents applications of Digital Voices, and section 6 concludes the paper.

## 2. BIT RATE OF HUMAN SPEECH REVISITED

Figure 1 shows a model of the speech-production / speech-perception process as suggested by Rabiner [6]. The bit rates of 50 bps associated with the highest level of the process can be traced to studies made in the 1950's; a summary of those studies can be found in [1].

The phonetic transcription of speech motivates a simple computation that justifies those low bit rates. Take, for example, English. In a simplified model, there are about 40 phonemes in the English language. Assuming an equal probability of occurrence of the phonemes, a straightfor-ward coding leads to *log2(40)=5.3* bits per phoneme. In conversational speech there is an average of 10 phonemes per second. That results in 53 bps. A coding scheme that takes into account the relative frequencies of utterance of the phonemes results in a smaller bit rate.

In the lowest level of the process, the acoustic waveform, the information content is about 1,000 times higher. The reasons for this discrepancy in information content from the highest to the lowest level are still not fully understood.

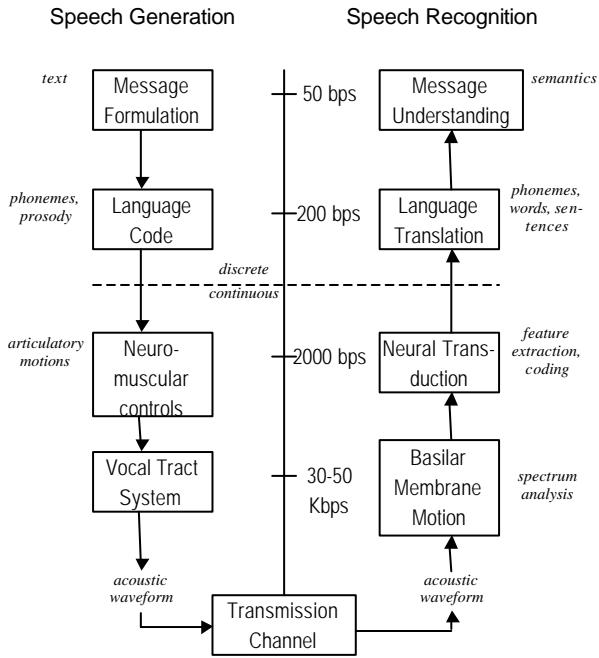

**Figure 1** Model of speech-production / speech-perception process (after Rabiner [6]).

## 3. DIGITAL VOICES AND THEIR CHANNELS

The goal of our digital voices is to transmit arbitrary information from one machine to one or more machines nearby using audible sound in air, similarly to what humans do through speech. But instead of aiming for speech-like sounds, we started by analyzing common modulation techniques used in digital communications such as Amplitude-Shift Key (ASK) and Frequency-Shift Key (FSK). We observed that some parameters of the modulation that have an impact in the data rate, the error probability and the computational overhead at the receiver also have a tremendous impact in the quality of the sound as perceived by humans. Variations of those parameters allow us to obtain messages ranging from modem noises to music. We designed combinations of those modulation schemes, in which we include a language that sounds like R2D2 – the robot in the movie Star Wars.

More recently, we have explored amplitude, frequency and phase modulations of fixed sets of sounds, inspired by the concept of phonemes in speech. In these we include a voice that sounds like a cricket. We've also experimented with English phonemes – an experimentation by no means intended to produce speech synthesis, but simply to study the use of signals similar to human speech.

It is very clear that the communication channel has a tremendous impact in the communication itself and in the signal design. The human communication channel in which speech exists has important differences from the communication channel we use for the digital voices. But there are also some similarities.

The human communication channel consists of the human vocal system, the air (medium) and the human hearing system. The vocal system imposes constraints on the kinds of sounds that can be produced and ultimately determines the sounds of speech. For example, humans will never be able to sound like crickets. The air defines some properties of the communication such as the fact that it's broadcasted, how far speech can propagate and the fact that it is a noisy medium. The hearing system imposes some more constraints on the sounds that can be used in communication and on the perception of sounds in general. For example, humans can't distinguish the difference between 1000Hz and 1005Hz. There is also a non-uniform filter of the frequencies in the hearing range.

The communication channels in Digital Voices consist of a generic sound production system implemented in software (coder/modulator), ordinary speakers, the air (medium), ordinary microphones and a generic sound perception system implemented in software (demodulator/decoder). The coders and decoders handle digital representations of sound and allow us to pursue the experiments described in this paper. For example, we can synthesize and perceive sounds of a violin, a cricket and even speech-like sounds. We can also distinguish accurately a tone of 1000Hz from another tone of 1005Hz. The speakers and the microphones, however, impose constraints on the kinds of sounds and modulations we can use. Tests we made on this kind of hardware have shown that it introduces non-uniform frequency attenuation and, in some cases, specific frequencies are attenuated to almost zero.

## 4. DESIGNS AND OBSERVATIONS

In this section, we describe a meaningful subset of our digital voices. Case 1 contains examples of standard modulation techniques that are the basis of several of our digital voices. Case 2 is an example of a well-known artificial "language" that is a sequence of child-friendly noises. Case 3 is an example of our work with natural non-speech sounds. Case 4 describes encodings of URLs that illustrate the role that the information structure can have in the communication. We make quantitative and qualitative tion. We make quantitative and qualitative observations about data rate, computational overhead, psychoacoustic effects and semantics.

Case 1: ASK and FSK. In ASK modulation, the message is encoded in the signal amplitude. In FSK modulation, the message is encoded in the frequency of the signal (see [5] for details).

We designed several 8-frequency B-ASK schemes. For a symbol duration $T=20ms$, the data rate is 400 bps. For $T=100ms$, the data rate is 80 bps. For $T=20ms$, the sound is similar to sounds of grasshoppers. For $T=100ms$ it sounds like a piece of music played by several instruments (soprano flutes, maybe). Another design uses 128-frequency ASK. In this case, all frequencies were harmonics of 70 Hz, starting at 700 Hz. For $T=100ms$, the data rate is 1280 bps. The sound is quite different from the previous design: it sounds like an electronic drum beat. In one of our FSK designs, 256 frequencies are separated by intervals of 20 Hz, starting at 1000 Hz, to transmit an 8-bit value in each $T=20ms$. The bit rate is 400 bps. The sound resembles one grasshopper. When we increase the duration of the symbol interval to $T=100ms$ it sounds like a bird.

Although the data rates for 8-frequency ASK and 256-ary FSK are the same for the same values of $T$, the resulting sounds are quite different (see [5]). In any case, humans can decode none of these schemes. These schemes show that even with in a slow medium like sound in air it is possible to transmit information at much higher bit rates and with a much lower computational overhead at the receiver than what speech does. Clearly humans are far from reaching the capacity of aerial channels.

Case 2: R2D2. We have designed a communication system sounding like R2D2 that can send ASCII text messages. Studying the original noises made by R2D2 in the movies, we've identified three major kinds of sounds: beeps, chirps and grunts. We noticed that, in the movies, R2D2 beeps much more than it chirps or grunts. We've also noticed that grunts are usually associated with a particular emotional state of the robot, i.e. bad mood or scorn, but we have intentionally ignored emotional states.

We wanted to get a reasonable distribution of these three modes for text messages, therefore we've established the following code: (1) letters, a-z, are mapped into beeps of 26 different frequencies (FSK modulation); (2) the space, and 3 other special characters such as '.', are mapped into 4 different chirps; and (3) numbers, 0-9, are mapped into 10 different grunts, which result from combinations of 4 frequencies (ASK modulation). Overall, we have 40 different symbols. To comply with the observed noises of the original R2D2, beeps, chirps and grunts have slightly different durations, namely 100 ms, 250 ms and 200 ms respectively.

The R2D2 digital voice can transmit an average sentence (i.e. 12 words, 5 characters per word) in about 9 seconds. This means the average transmission bit rate is about 35 bps, which, according to the model in Figure 1, is only slightly lower than human speech. The computational overhead at the receiver is, again, much lower than the computational overhead of speech recognizers.

Case 3: Cricket. Departing from the previous cases, we experimented with animal sounds, to understand the information capabilities of those signals. If animals were to transmit arbitrary information through their songs and sounds, where could that information be? [1]

Here we describe cricket songs. Crickets have a characteristic type of song consisting of trains of three fast beeps. Figure 2 shows an excerpt of a real cricket song.

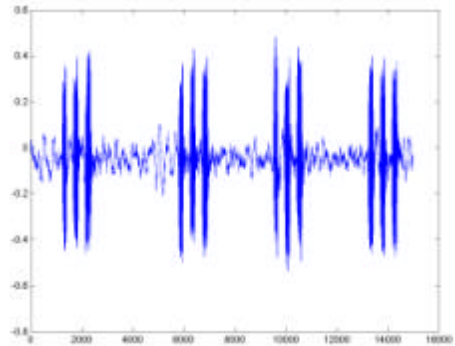

**Figure 2.** Excerpt from a real cricket song.

In order to preserve the nature of the cricket songs, we modulated information both in the phase of the triadic beeps and in the amplitude of the beeps. We have defined 32 acoustic symbols (or phonemes). This system can transmit information at about 22 bps. The receiver is still simple, but it's more complex than the receivers for the previous designs. In playing all these different digital voices to human listeners, the cricket songs were found to be a favorite sound.

Case 4: Poly-semantic codes. Most digital communication systems in use today were designed with the goal of transmitting large amounts of unstructured data, such as images, as fast as possible.

But, very often, data follows rules. Take, for example, URLs, which, on a first look, can be seen as arbitrary sequences of ASCII characters. There are several kinds of structures underlying URLs. First, they include well-known sub-strings that occur often, like, for example, "http://" or "mailto:" or ".com". This knowledge enables us to use appropriate source coding to reduce the number of bits to be sent. Second, URLs comply with pre-defined grammars. The information that comes from the grammar can be used as a code to transmit URLs. For example, "mailto" and "http" URLs can be transmitted in two different bands of the audible spectrum, say B-ASK at 1000Hz and 2000Hz respectively. Grammar-based codings expose meta-information associated with the data.

---

[1] From what is known in bioacoustics, animal songs are aids for functions of survival and reproduction. Although there is no reason to believe that animals communicate arbitrary information, as humans do, for the artificial communication systems of Digital Voices, this is an interesting question.

This opens up a design space for digital voices that consists in sounds that expose different aspects of the information content for different receivers. This is what we call poly-semantic codes. One aspect might be the grammar, but there are other aspects such as the intended recipient or the urgency of the message, if the messages contain such pieces of information. Some applications may benefit from digital voices that convey some bits of information, but not all, to the humans listening to the communicating machines. This idea builds on work done in auditory displays [4].

Speech, at least for English and Roman-based languages, doesn't seem to have any grammar-based codings, i.e. sounds that are specific to, for example, verbs, although, as digital voices shows, it could. But speech often conveys information that is coded using an orthogonal code to that of the verbal language. For example, the pace, pitch and volume at which the sentences are uttered convey information that is usually omitted from the verbal message, even though it could be included in it, and that is meaningful even for those who don't understand the language. The non-verbal information of speech has been particularly challenging to model, and, for that reason, has been largely ignored in speech recognition.

## 5. APPLICATIONS

We believe Digital Voices are useful beyond serving as comparison to human speech. Sound in air is attractive for applications that do not require high bit rates and for which it is expensive to extend the hardware infrastructure with radio or infrared transmitters. It is also attractive for applications that benefit from human awareness of the communication.

Some examples of those applications are: toys; broadcasting information through the sound of TV and radio that can be picked up by devices at home or in the car; transferring names and phone numbers between cell phones; transferring business cards between PDAs; and broadcasting location-dependent information from rooms into PDAs and laptops (context-aware computing). We are also looking at applications for security in wireless networks, using sound as a location-limited channel.

In all these applications, we might imagine using synthesized speech, rather than digital voices, for the devices to communicate with each other. But speech is not necessarily the best means for representing the information we might want to transmit. For example, using speech to transmit URLs is slower, more computationally intensive and more error-prone than it needs to be. An R2D2 kind of voice is much simpler to implement, and much easier to decode accurately.

## 6. SUMMARY

We have presented a framework and a method that can be used for analyzing the highest level of the speech process. The framework defines objective functions of the communication such as data rate, computational overhead, psychoacoustic effects and semantics. We have used this framework to compare the verbal information of speech with other low bit rate acoustic communication systems implemented as air modems. In the future, the framework can be extended with other objective functions, and can also be used to analyze non-verbal information such as urgency or emotional state.

**Acknowledgments**. Patricio de la Cuadra from Stanford University, and Ling Bao from MIT implemented the cricket voice and the R2D2 voice during their summer internship at PARC.